# Switchable Rashba Anisotropy in Two-dimensional Hybrid Organic-Inorganic Perovskite by Hybrid Improper Ferroelectricity

Fei Wang, Heng Gao, Coen de Graaf*, Josep M. Poblet,

Branton J. Campbell & Alessandro Stroppa*

## Abstract

Two-dimensional (2D) hybrid organic-inorganic perovskites (HOIPs) are introducing new directions in the 2D materials landscape. The coexistence of ferroelectricity and spin-orbit interactions play a key role in their optoelectronic properties. We perform a detailed study on a recently synthesized ferroelectric 2D-HOIP, (AMP)PbI$_4$ (AMP = 4-aminomethyl-piperidinium). The calculated polarization and Rashba parameter are in excellent agreement with experimental values. We report a striking new effect, *i.e.,* an extraordinarily large Rashba anisotropy that is tunable by ferroelectric polarization: as polarization is reversed, not only the spin texture chirality is inverted, but also the major and minor axes of the Rashba anisotropy ellipse in *k*-space are interchanged – a pseudo rotation. A $k \cdot p$ model Hamiltonian and symmetry-mode analysis reveal a quadrilinear coupling between the cation-rotation modes responsible for the Rashba ellipse pseudo-rotation, the framework rotation, and the polarization. These findings may provide new avenues for spin-optoelectronic devices such as spin valves or spin FETs.

## Introduction

In recent years, there has been an intense research effort surrounding the conversion of solar energy into electric power through photovoltaic cells.[1-3] Hybrid organic-inorganic perovskites (HOIPs) with composition ABX$_3$ (A = organic cation, such as [CH$_3$NH$_3$]$^+$; B = divalent metal cation, such as Sn$^{2+}$, Pb$^{2+}$; X = halogen anion) are proposed as a new generation of materials for solar cells and light-emitting diodes.[4-8] The presence of new structural degrees of freedom from the inclusion of organic cations into a BX$_3$ inorganic framework enables structural distortions and columnar shifts which are otherwise forbidden in standard inorganic perovskites. Furthermore, ferroelectricity (FE) originating from structural symmetry breaking is a potentially critical phenomenon in such soft and flexible photovoltaic materials, where the electric field plays an important role in promoting electron-hole pair separation and suppressing charge recombination, inherently breaking the Shockley-Queisser limit.[9-13] So the FE properties of perovskites have attracted broad attention, though their presence and influence in perovskite solar cells are still a matter of debate.[14-19]

A strong spin-orbit coupling (SOC) involving the heavy Pb atom and its surrounding halogens, combined with the absence of an inversion symmetry in the crystal structure, can lead to a Rashba effect and thus impact photovoltaic performance.[20-23] Non-centrosymmetry provides a spin-degenerate band with the possibility to split two reversely spin-polarized states, and the equation of $E_\pm(k) = \frac{\hbar^2 k^2}{2m} \pm \alpha_R |k|$ can be used to clarify the dispersion relation of electrons (or/and holes), in which $\alpha_R$ is the Rashba splitting parameter.[24,25] The coexistence of ferroelectricity and a Rashba effect is believed to mediate interesting effects, such as the switching of spin-texture chirality with an electric-polarization reversal.[26] These effects are mainly studied in the context of inorganic materials; only a few hybrid-material examples have been explored so far.[24,27,28]

In the HOIPs family of compounds, 3D systems have received far more attention than 2D systems, though the 2D counterparts possess exceptional chemical stability and novel and less-explored photovoltaic capabilities.[4,29] Although the 2D systems are more likely to support an FE polarization,[30-33] the experimental evidence of the coexistence of FE and Rashba effects in 2D-HOIPs is very rare. Very recently, Sum *et al*. demonstrated the presence of both ferroelectricity and a Rashba effect in a 2D Dion-Jacobson (DJ) phase HOIP with formula (AMP)PbI$_4$, where AMP is the divalent 4-aminomethyl-piperidinium cation (Figure 1, FE state). Its measured value of spontaneous polarization ($P_s$) is 9.8 $\mu$C/cm$^2$, which is very high amongst HOIPs[34,35], and more comparable to those of conventional inorganic perovskites such as BaTiO$_3$. A measured Rashba splitting energy ($E_R$) of 85 meV and Rashba coefficient ($\alpha_R$) of 2.60 eV·Å suggest real promise for spintronic applications. This exciting new experimental evidence now motivates a theoretical investigation into the microscopic mechanism of ferroelectricity and its interplay with spin-orbit coupling in the 2D-HOIPs.

Starting with this aim in mind, we perform a detailed density functional theory (DFT) analysis of (AMP)PbI$_4$, focusing on estimates of the ferroelectric and spin-orbit properties of the band structure. While the estimate of the Rashba splitting is now a routine work, we focus here on the corresponding anisotropy, which is far much less studied. Our estimate of $P_s$ and $\alpha_R$ are 10.72 $\mu$C/cm$^2$ and 2.39 eV·Å, in excellent agreement with the experimental 9.80 $\mu$C/cm$^2$ and 2.60 eV·Å.

In addition, the $E_R$ of spin-polarized bands is calculated to be 16 meV, which is of the same order of magnitude as the experimental 85 meV value. In addition to a spin-texture that can be switched via ferroelectric polarization reversal, we further observe a sizeable anisotropy in the Rashba spin splitting. Remarkably, we show that the axis of major anisotropy can be switched between two orthogonal directions by reversing the ferroelectric polarization from $+P$ to $-P$. Reversing the FE polarization induces an apparent 90° 'rotation' of the anisotropy ellipse, *i.e.,* a pseudo-rotation. This new effect has never been highlighted in the context of usual spin-texture tunability under polarization reversal and may suggest new avenues for spin-optoelectronic devices based on hybrid organic-inorganic perovskites.

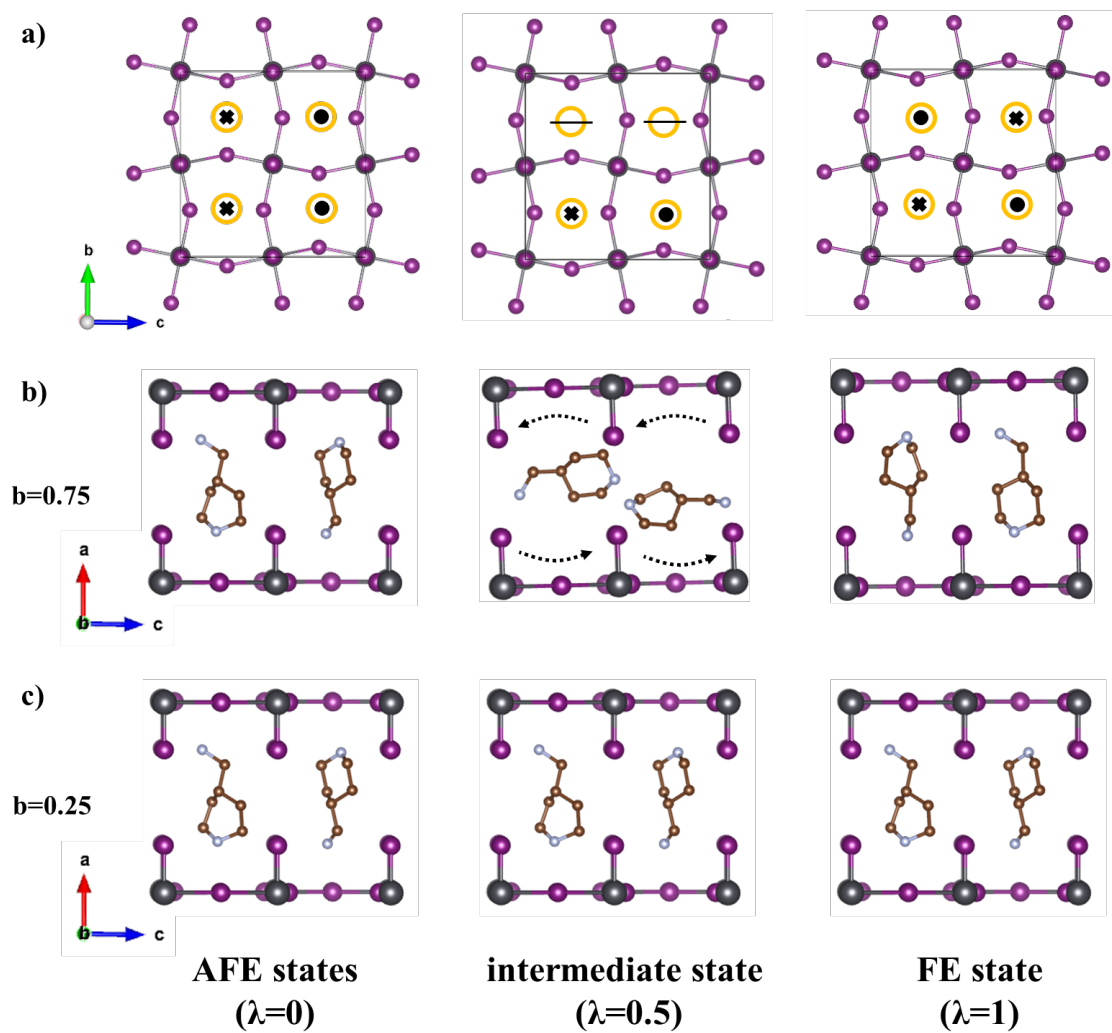

**Figure 1 | Top (a) and side (b and c) views of the of AFE and FE (+*P*) phases**, with purple Pb, gray I, brown C and blue N. AMP-cation hydrogen atoms are omitted for clarity. The black dot or cross in the yellow circle of **(a)** represent the AMP-cation orientation as having the aminomethyl group of the molecule along the up (+*a*) or down (−*a*) axis, respectively.

## Results

**Structural analysis:** We start our simulations with the experimental ferroelectric phase of (AMP)PbI$_4$ refined at $T$ = 298 K having space group $Pc$ (#7, $C_s^2$) with monoclinic *b*-axis in Ref.[36] Two nearby AMP$^{2+}$ cations located in the space between [PbI$_4$]$^{2-}$ layers have alternating orientations when adjacent along the *b* or *c* axes, but common orientations when adjacent along diagonal $b + c$ or $b - c$ directions, as shown in Figure 1 (FE state), so that the structure lacks a center of inversion, thus paving the way for a possible ferroelectric polarization. It is interesting to note that the cation's center of mass is displaced somewhat from the ideal midpoint between two adjacent $[PbI_4]^{2-}$ layers so as to be slightly closer to one layer or the other. We perform quantum chemical analyses of the isolated $AMP^{2+}$ cation using both the Hirshfield[37] and Natural population methods[38] implemented in the Gaussian16 software.[39] Both methods define an electric dipole moment of the molecule pointing approximately towards the aminomethyl group from the center of the AMP cation, as shown in Supplementary Figure 2. Although the long-range ordering of AMP-cation orientations should be associated with a strong FE polarization along the *a* axis at the first glance, this alone is not sufficient for understanding the overall polarization of a hybrid system, which typically has several different contributions to the polarization, as explained in details in Ref. 32. According to the modern theory of polarization,[40] we introduce a centrosymmetric anti-ferroelectric (AFE) phase by enforcing the existence of an inversion point, and define a suitable path connecting the AFE and FE state via atomic rotations or/and displacements in order to estimate the polarization by first principles calculations. The AFE model is built from the FE model by simultaneously rotating two $AMP^{2+}$ cations adjacent along the *b*-axis by 180° so that the diagonal AMP cations are related by inversion symmetry. The $[PbI_4]^{2-}$ framework is also properly centrosymmetrized to ensure that the AFE structure (both framework and AMP cations) possesses an inversion symmetry point at (0.5, 0.5, 0.5), as shown in Figure 1 (AFE state). The resulting space group symmetry is $P2/m$ (#10, $C_{2h}^1$) with monoclinic *b*-axis in the same unit cell.

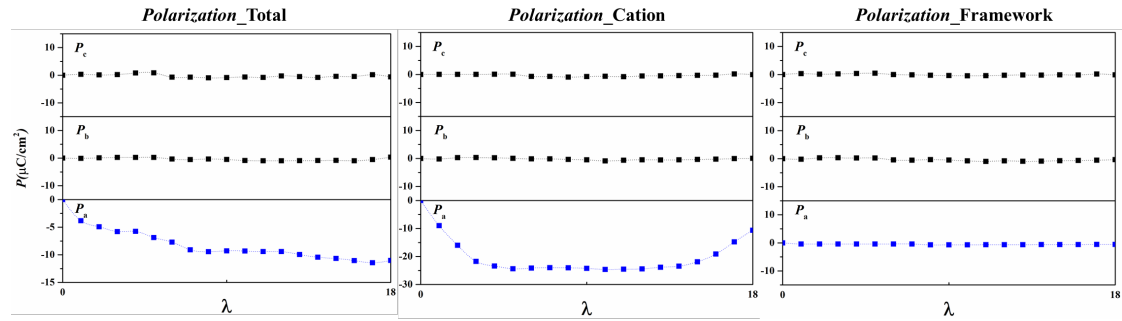

**Figure 2 |** The FE polarization along the *a*, *b*, and *c* axes of (AMP)PbI$_4$ as a function of the parameter $\lambda$, which represents the normalized $AMP^{2+}$ cation rotations and $[PbI_4]^{2-}$ framework displacements.

**Ferroelectric properties:** We introduce the normalized variable $\lambda$ to simultaneously parameterize both the rotations of the AMP cations and the atomic displacements of the $[PbI_4]^{2-}$ framework from AFE ($\lambda = 0$) to FE ($\lambda = 1$) states. Therefore $\lambda$ represents the amplitude of the combined roto-displacive distortion connecting the prototype AFE-phase structure and the real FE-phase structure. The intermediate structures along the path ($0 < \lambda < 1$) are only introduced in order to monitor the accidental introduction of a quantum of polarization and to ensure a continuous change of the polarization along the path itself. In practice, $\lambda$ has been varied in finite increments, so that continuous AMP-cation rotations are approximated by 18 steps of 10° each from 0° (AFE state) to 180° (FE state), together with incremental linear displacements of the $[PbI_4]^{2-}$ framework.

In order to understand the origin of the polarization, the separate contributions of the organic $AMP^{2+}$ cations and the inorganic $[PbI_4]^{2-}$ framework to the total polarization are considered in three different ways. Firstly, we rotate AMP cations while keeping the $[PbI_4]^{2-}$ framework centrosymmetric. In this way, we obtain the contribution coming from the organic groups ($P_{cation}$), *i.e.,* a rotational contribution to the polarization. Secondly, the displacive distortions of $[PbI_4]^{2-}$ framework are considered while fixing AMP cations in a centrosymmetric configuration. In this way, the polarization only originates from the inorganic framework ($P_{frame}$), *i.e.,* a displacive contribution to the total polarization. Finally, both of AMP-cation rotations and $[PbI_4]^{2-}$ layer displacements are simultaneously activated in order to recover the total polarization ($P_{total}$), *i.e.,* the full roto-displacive contribution to the polarization. In this way, we are able to disentangle the different contributions to the polarization and to observe the couplings between them. For each case, we consider the contribution from the ionic and electronic subsystems to the resulting polarization.

The results of the polarization calculations are presented in Figure 2. In the AFE state, the net polarization is exactly zero, as expected. As $\lambda$ gradually increases towards 1, the asymmetry with respect to the centric phase and the magnitude of the total polarization both increase steadily along the *a*-axis but remain near zero for the *b* and *c*-axis components. For ($\lambda = 1$), we extract the estimated value of $P_{total}$, which is equal to 10.72 $\mu C/cm^2$, in very good agreement with the experimental value of 9.80 $\mu C/cm^2$. From our analysis, the contributions from organic cations and inorganic framework are 10.34 $\mu C/cm^2$ and 0.59 $\mu C/cm^2$ respectively. Clearly, the AMP cations dominate the total polarization, so that the framework contribution appears negligible in comparison. An applied electric field then mainly rotates the AMP cations while displacing the $[PbI_4]^{2-}$ framework only a little. Because the switchable FE polarization is primarily due to the long range orientational order of the organic cations, the true AFE state lying between the +*P* and –*P* states could be consistent with either ordered or disordered but balanced arrangements of AMP-cation orientations, thus supporting a null electric polarization. Indeed, the AFE ordered state of the crystal represents only a computational reference phase to evaluate the final polarization, and in principle may be not uniquely defined. In the present case, we have fixed a possible AFE ordering, but, according to modern theory of polarization, the final polarization does not depend on the particular path considered for $0 \leq \lambda < 1$.[41]

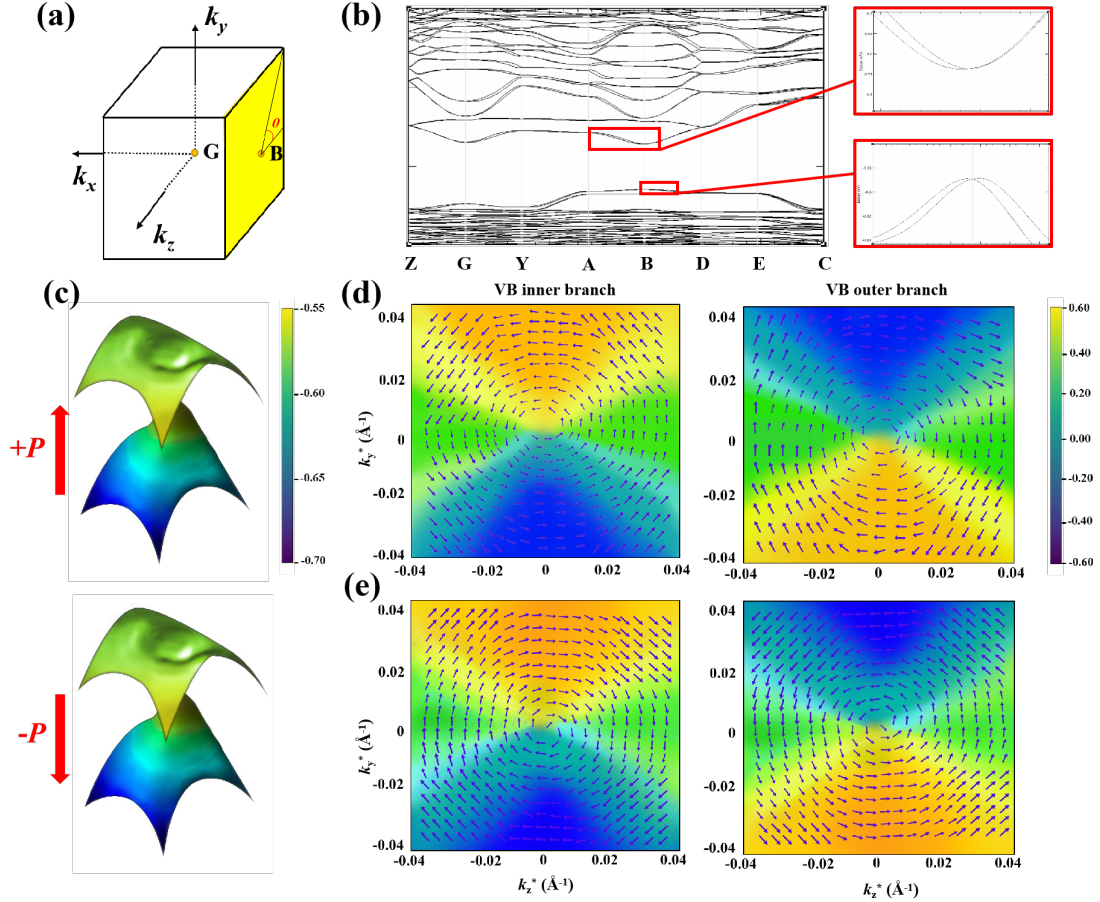

**Figure 3 | The band structure of FE state. (a)** The high-symmetry points of BZ. **(b)** Dispersed band structures along the whole BZ; the spin-split bands of the VBM (bottom) and CBM (top) were magnified in the red frames for emphasis. **(c)** 3D plot of band dispersion for valence bands (VBs) around B point. The color evolution from blue to yellow represents the VB eigenvalues. **(d, e)** The spin-textures projected on the 2D plane perpendicular to the FE polarization (*P*). The color code represents the $s_x$ spin component.

**Electronic structure and Rashba SOC effect:** We now examine the electronic structure and Rashba SOC effect of ferroelectric (AMP)PbI$_4$. The Brillouin Zone (BZ) and corresponding Rashba spin-polarized bands of the FE structure are shown in Figure 3a and 3b. The valence band maximum (VBM), conduction band minimum (CBM) and related band gap are localized in *k*-space around the high-symmetry B point. The VBM and CBM are strongly spin-split due to the spin-orbit interaction and are also shifted away from the B point in *k*-space. As usual, the "momentum offset" ($k_R$) corresponds to the distance between the apex of the splitting band and the high-symmetry point in *k*-space, while $E_R$ corresponds to the energy difference between them. In our case, $k_R$ and $E_R$ are estimated as 0.0133 Å$^{-1}$ and 16 meV for the VBM and as 0.0132 Å$^{-1}$ and 12 meV for the CBM, which shows rather good agreement with the respective experimental values of 0.067 Å$^{-1}$ and 85 meV. Indeed, according to the definition $α_R = 2E_R/k_R$, $α_R$ is calculated to be 2.39 eV·Å for the VBM, which agrees with the experimental 2.60 eV·Å very well. It must be noted that the $k_R$ values for the VBM and CBM are very similar, indicating that they are located essentially at the same point in *k*-space to form a "direct" band gap, which is beneficial for electronic transitions in photovoltaic-cell applications.

We show the band dispersion of the spin-split bands, *i.e.,* a 3D plot of the VBs around the high-symmetry B point, in Figure 3c. The corresponding spin-textures of the spin-split bands are projected onto the 2D plane perpendicular to the FE polarization (*P*) in Figure 3d. In the inner and outer bands, the sense of rotation (left-handed or right-handed) of the spin-textures are opposite due to the spin-orbit interaction. As reported in Refs.[42,43], the polarization reversal resulting from the application of an external electric field should be able to switch the spin textures. Indeed, we also observe that the spin textures completely switch from left- to right-handed sense of rotation as the polarization goes from –*P* to +*P*, as shown in Figure 3d-3e.

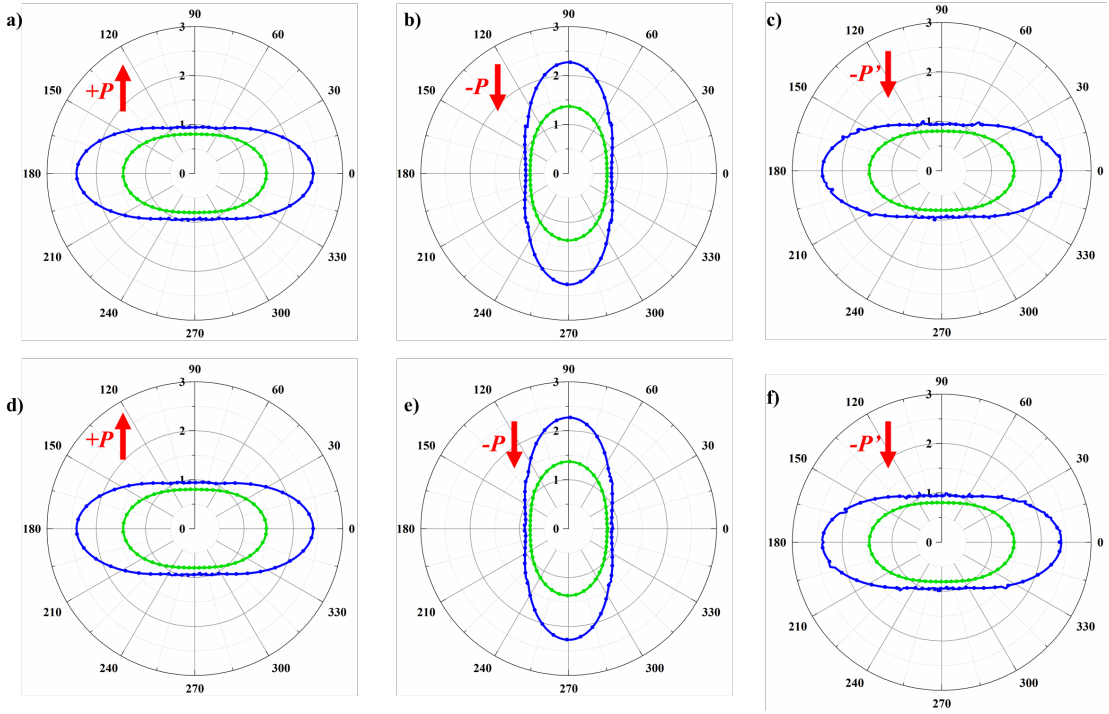

**Figure 4 | Polar plots of variable $\alpha_R$ for Rashba spin-split bands obtained along different directions in the plane perpendicular to the FE polarization.** The plots of **a-c** are acquired at the B point while those of **d-e** are at the G point. The blue curves correspond to the $\alpha_R$ obtained from the VBs while the green curve corresponds to those from the CBs. The red arrows indicate the direction of the polarization, and –*P*' indicates an alternative –*P* structure wherein each AMP molecule rotates 180° relative to the +*P* structure around the axis perpendicular to its maximum-electron-density plane rather than around the *b* axis. However, this –*P*' is energetically unfavorable as discussed in text.

While most studies on spin-orbit related properties estimate the Rashba parameter for a given direction (in the plane perpendicular to the polarization) and visualize the spin-texture and its switching properties, here we also focus on the anisotropy of the Rashba parameter and its relationship to the switching of the ferroelectric polarization. To probe this anisotropy, the band structure was calculated on a fine grid of points in the vicinity of the reciprocal-space B point within the 2D plane perpendicular to the FE polarization (the yellow $k_y k_z$ plane in Figure 3a). The calculated $\alpha_R$ parameters are summarized in Figure 4a, where the blue curves correspond to the $\alpha_R$ obtained from the VBs and the green curves corresponds to those from the CBs. It can be seen that $\alpha_R$ parameters show a marked anisotropy, both at VBs and CBs. The maximum $\alpha_R$ of the VB at the B point (the blue curve in Figure 4a) is 2.39 eV·Å, located on the $k_z$-axis, whereas the minimum is 0.94 eV·Å, located on the $k_y$-axis, which indicates that $\alpha_R$ is not a constant but rather a function of the angle $\theta$ with respect to the negative $k_z$-axis (in the yellow plane in Figure 3a).

Remarkably, we find that when the FE polarization is switched from $+P$ to $-P$, the major and minor axis of the Rashba anisotropy are exchanged as shown in Figure 4b, so that the maximum $\alpha_R$ value of the VB (2.22 eV·Å) shifts from the $k_z$-axis (b-axis) to the $k_y$-axis (c-axis), and the minimum $\alpha_R$ value (0.87 eV·Å) shifts the opposite way. In other words, while the polarization switches from $+P$ to $-P$, the major axis $\alpha_R$ value decreases, while the minor axis $\alpha_R$ value increases. If we compare the $+P$ and $-P$ anisotropy ellipses, the net effect is an apparent 90° rotation of the anisotropy ellipse, i.e., a pseudo-rotation. To the best of our knowledge, this is the first time that a polarization-switchable Rashba anisotropy has been observed, which is also connected to a very large anisotropy.

We propose a structural mechanism for this highly unusual polarization-switchable pseudo rotation of the Rashba anisotropy. As the AMP-cation orientations evolve between the $+P$ and $-P$ states, the roughly 180° rotation (approximately around the b-axis) of each cation not only switches the aminomethyl group between the $+a$ and $-a$ sides of the cation, but also rotates its maximum-electron-density plane (MEDP) by about 90° in the bc plane, which is dual to the $k_y k_z$ plane of the BZ (can be seen in Supplementary Figure 3). To verify our hypothesis that this polarization-reversal-induced 90° rotation of the AMP cation's MEDP is responsible for the rotation of the Rashba-anisotropy, we construct an alternative $-P'$ structure from the $+P$ structure by rotating each AMP cations by 180° around the axis perpendicular to its MEDP, so that its aminomethyl group is still moved between the $+a$ and $-a$ sides, but without allowing the MEDP to rotate in the bc plane. When the structure is thus evolved, we see no rotation of the Rashba anisotropy in its band structures (Figure 4c and 4f). Furthermore, compared to the isoenergetic $+P$ and normal $-P$ structures, the alternative $-P'$ structure has a 0.48 eV higher energy than $+P$, and is separated from $+P$ by an energy barrier of greater than 1.20 eV, demonstrating that the $-P'$ structure would be very difficult to obtain at room temperature.

We derive the $k \cdot p$ model near the B point using point group C$_{2v}$. The effective Hamiltonian up to second order reads as $H(k) = \varepsilon_0 + \frac{\hbar^2 k_y^2}{2m_y} + \frac{\hbar^2 k_z^2}{2m_z} + \alpha_z k_z \sigma_y + \alpha_y k_y \sigma_x$, Here $m_y$ and $m_z$ are the effective masses along the $k_y$ and $k_z$ directions near the B point, and $\sigma_x$ and $\sigma_y$ are Pauli matrices. Then $E(k) = \varepsilon_0 + \frac{\hbar^2 k_y^2}{2m_y} + \frac{\hbar^2 k_z^2}{2m_z} \pm \sqrt{\alpha_z^2 k_z^2 + \alpha_y^2 k_y^2}$, or $E(k_r, \theta) = \varepsilon_0 + (a_1 + a_2 \cos(2\theta))k_r^2 \pm \sqrt{\alpha_1 + \alpha_2 \cos(2\theta)}|k_r|$ in polar coordinates. Here $k_r$ is defined as the distance of $k$ from the B point in the $k_{yz}$ plane, and $\theta$ is the angle between $\vec{k}_r$ and $\vec{k}_z$. Substituting $k_z = k_r \cos(\theta)$, $k_y = k_r \sin(\theta)$, $a_1 = \frac{1}{2}\left(\frac{\hbar^2}{2m_y} + \frac{\hbar^2}{2m_z}\right)$, $a_2 = \frac{1}{2}\left(\frac{\hbar^2}{2m_y} - \frac{\hbar^2}{2m_z}\right)$, $\alpha_1 = \frac{1}{2}(\alpha_y^2 + \alpha_z^2)$, and $\alpha_2 = \frac{1}{2}(\alpha_z^2 - \alpha_y^2)$, where $\alpha_1$ = 3.30 (eV·Å)² and $\alpha_2$ = 2.41 (eV·Å)² in the $+P$ state, the Rashba parameter is then

$$\alpha_R^+ = \sqrt{\alpha_1 + \alpha_2 \cos(2\theta)} = \sqrt{3.30 + 2.41 \cos(2\theta)}.$$

Because the rotation of the AMP cation in the *bc* plane effects a 90° rotation of the crystal fields at the cation site, $a_1$ and $\alpha_1$ keep the same but $a_2$ and $\alpha_2$ change sign. The anisotropy of Rashba parameter at –*P* state is:

$$\alpha_R^- = \sqrt{\alpha_1 - \alpha_2 \cos(2\theta)} = \sqrt{\alpha_1 + \alpha_2 \cos\left(2\left(\theta + \frac{\pi}{2}\right)\right)} = \sqrt{3.30 + 2.41 \cos\left(2\left(\theta + \frac{\pi}{2}\right)\right)},$$

which is rotated by 90° relative to $\alpha_R^+$. The proposed $k \cdot p$ model is thus nicely consistent with our DFT results in terms of a 90° pseudo-rotation.

We introduce parameter $\mu = \left|\frac{\alpha_R(k_z)}{\alpha_R(k_y)} - 1\right|$ to quantify the strength of the anisotropy where $\alpha_R(k_z)$ and $\alpha_R(k_y)$ are the numerical values of the Rashba parameter along the $k_z$ and $k_y$ axes of the BZ, respectively. The larger the value of $\mu$, the more pronounced the anisotropy of $\alpha_R$ is. In the case of the –*P* and +*P* polarizations, $\mu$ has quite similar values of 1.7 and 1.6, respectively, reflecting the robustness of the Rashba anisotropy in this system.

It is interesting to note that the pseudo-rotation of the Rashba anisotropy induced by polarization reversal occurs not only at a single point in *k*-space, but at several points where the Rashba splitting is effective. Our observations of the Rashba anisotropy in the VB and CB near both the B and G points of the BZ are all qualitatively consistent with the description and explanation of Figure 4d-f above. At both of these points, the pseudo-rotation of the anisotropy ellipse occurs when reversing the polarization perpendicular to the framework layers.

**Symmetry mode analysis:**

The +*P* and –*P* FE structures can be viewed as distorted "children" of an idealized "parent" structure, which allows us to characterize their symmetry-breaking structural variables (*i.e.*, symmetry modes) in terms of the irreducible representations (irreps) of the parent space-group symmetry. We construct this idealized parent from the centrosymmetric AFE structure in the left panel of Figure. 1a by regularizing the PbI$_6$ octahedra and zeroing the PbI$_6$ octahedral rotations in order to straighten out the framework, and by replacing each AMP cation with a polar vector (attached to a dummy atom at the molecular centroid location) that points along the appropriate ±*a* direction, resulting in centrosymmetric space group *Pcmm* (#51, $D_{2h}^5$) on an orthorhombic (*a*, *b*/2, *c*) unit-cell basis with zero origin shift relative to the child. We also idealize the FE children by replacing each AMP cation with an appropriate polar vector in order to apply the symmetry mode analysis consistently.

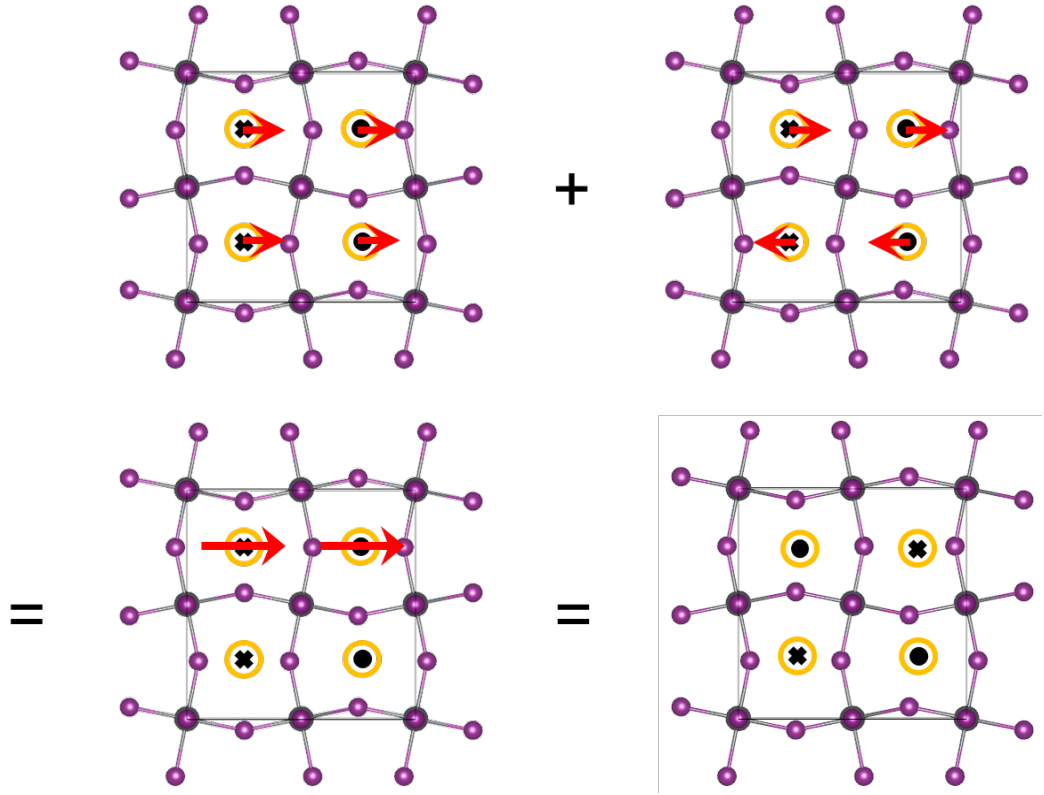

**Figure 5 | Molecular rotation modes.** Short (90°) and long (180°) red arrows indicate right-handed molecular rotations relative to the AFE configuration. Summing the FR (upper left) and AFR (upper right) patterns of 90° rotations causes only two of four molecules to rotate by 180° (lower left), which is precisely the FE structure (lower right).

The key order parameters of the idealized parent structure that characterize our FE child structures are (1) the PbI$_6$ octahedral framework rotation around the $a$-axis, which belongs to parent irrep $Y_3^+$, (2) the ferroelectric displacements along the $a$-axis, which belong to FE parent irrep $\Gamma_2^-$, and (3) the large AMP-cation rotations around the $b$-axis, which are achieved through the cooperative action of two parent irreps, ferrorotational (FR) irrep $\Gamma_4^+$ and anti-ferrorotational (AFR) irrep $Y_1^-$. Using the INVARIANTS software package, we find that these four primary order parameters provide a quadralinear invariant term in the free energy expansion, so that simultaneously invoking the framework ($Y_3^+$) and the cation ($\Gamma_4^+$, $Y_1^-$) rotational modes breaks the inversion symmetry of parent, thus allowing them to couple to a secondary (*i.e.,* improper) ferroelectric moment ($\Gamma_2^-$).

Because the PbI$_4$ framework is only slightly disturbed in switching between the $+P$ and $-P$ states, it is convenient to treat the $Y_3^+$ framework rotation as a large preexisting structural feature in a slightly less-symmetric parent structure with space group *Pnmm* (#59, $D_{2h}^{13}$) and unit-cell basis and origin identical to those of the child. For this this less-symmetric parent structure, the FR and AFR cation rotations form a simple trilinear invariant with the ferroelectric moment.

The $+P$ cation arrangement is obtained from the parent via the superposition of a 90° FR motion ($\Gamma_4^+$) and a 90° AFR motion ($Y_1^-$), so that two of the AMP cations are rotated by a full 180° while the other two remain stationary (as shown in Figure. 1). The $-P$ cation arrangement is achieved instead by merely reversing the sense of the AFR contribution. Due to the unfavorable energy at the 90° midpoint of rotation, it makes no sense to view the FR and AFR rotations as separate processes; they must occur simultaneously and cooperatively, so that the $\Gamma_4^+$ and $Y_1^-$ order parameters are tightly coupled to have equal amplitudes. The product of two such large order parameters facilitates a strong trilinear coupling to the FE polarization. It remarkable that two centric modes, *i.e.*, $\Gamma_4^+$ and $Y_1^-$, combine together to give rise a hybrid mode, which in turn, breaks inversion symmetry and allows the polarization to arise in this system. Moreover, the coherent switching of both of them, *i.e.*, the switching of the hybrid mode, allows to switch the polarization giving rise to a hidden pseudo-rotation in *k*-space, as highlighted in this work with the switchable Rashba anisotropy ellipse. It is intriguing that the switchable Rashba anisotropy occurs coherently at different relevant point in *k*-space.

**Conclusions**

2D-HOIPs are emerging as a new class of photovoltaic materials. In this work, DFT methods have been used to study the ferroelectric properties and Rashba spin-orbit effect as well as the coupling between the FE polarization ($P_s$) and the anisotropy of the Rashba parameter ($\alpha_R$) for a 2D-HOIP, (AMP)PbI$_4$, where AMP is 4-(aminomethyl)piperidinium. We calculate $P_s$ and $\alpha_R$ to be 10.72 $\mu$C/cm$^2$ and 2.39 eV·Å, respectively, which are nicely consistent with the experimentally reported values of 9.8 $\mu$C/cm$^2$ and 2.60 eV·Å. To gain insight into the origin of the polarization, the contributions of the organic AMP$^{2+}$ cations (10.34 $\mu$C/cm$^2$) and inorganic [PbI$_4$]$^{2-}$ framework (0.59 $\mu$C/cm$^2$) have been disentangled, demonstrating that the AMP cations contribute almost all of the total polarization.

As previously reported for Rashba ferroelectrics, the spin-texture sense of rotation can be reversed by a reversal of the FE polarization direction along the polar axis.[40] However, in (AMP)PbI$_4$, we calculate a large and robust anisotropy in the Rashba parameter. Remarkably, we find that the major and minor axes of the ellipse can be exchanged under reversal of the electric polarization, causing a 90° pseudo-rotation of the Rashba anisotropy, which we have also been confirmed from a theoretical $k \cdot p$ model. The same effect is observed in both the VBs and CBs at multiple points in the Brillouin Zone where the Rashba splitting is significant. A structural mechanism for this novel effect is presented and explained in terms of a quadrilinear coupling of order parameters involving two large-amplitude AMP-cation rotation modes, a large octahedral framework rotation, and the ferroelectric polarization. To the best of our knowledge, this is the first report of a coupling of ferroelectric polarization and switchable highly-anisotropic Rashba spin-split bands. Spin-optoelectronic devices based on hybrid organic-inorganic trihalide perovskites like spin-LEDs have been recently discussed for the parent compound MAPbBr$_3$,[44] circularly polarized light detections,[45] spin-FETs. This strong coupling may provide a new probe of the spin degrees of freedom in photovoltaic materials and a new avenue for developing new spin-optoelectronic devices based on HOIPs materials.

**Methods**

**Computational Details**. The starting point for our calculations is the experimental crystallographic data reported in Ref. 33, where we relax all of the atoms of the ferroelectric structure until the Hellmann-Feynman forces are smaller than 0.001 eV/Å. The projector augmented wave (PAW)[46] method is used to solve Kohn-Sham equations with the PBE exchange-correlation functional,[47] as implemented in the Vienna *ab initio* simulation package (VASP). The energy cutoff for the plane wave expansion is set to 550 eV, and a 2×2×2 Monkhorst-Pack grid of *k*-point is used after systematic convergence tests. The Berry phase approach is used to evaluate the ferroelectric polarization through constructing a properly AFE reference state and a suitable path connecting FE with AFE states.[40,48,49] Details of the evaluation of the polarization are discussed in the results section. Spin-orbit interaction are self-consistently considered in all band structure calculations, but not in the calculations of polarization. Van Waals interactions are included in all of calculations.[50]

## Acknowledgments

This project has received funding from the European Union's Horizon 2020 research and innovation programme under the Marie Skłodowska-Curie grant agreement No. 713679 and from the Universitat Rovira i Virgili (URV). J.M.P. and C.d.G thank the Spanish Ministry of Science (grants CTQ2017-87269-P and CTQ2017-83566-P) and the Generalitat de Catalunya (grant 2017SGR629) for support. J.M.P. also thanks ICREA foundation for an ICREA ACADEMIA award. H.G. acknowledges support from the National Postdoctoral Program for Innovative Talents (No. BX20190361) and Guangdong Basic and Applied Basic Research Foundation (No. 2019A1515110965). B.J.C. acknowledges helpful conversations with Harold T. Stokes. B.J.C. and A.S. thank Z. V. Vardeny for useful discussions. A.S. acknowledges A. Cassinese, F. Chiarella and M. Barra for useful discussions.

# Author information

## Affiliations

Departament de Química Física i Inorganica, Universitat Rovira i Virgili, c/Marcel·lí Domingo 1, Tarragona 43007, Spain

Fei Wang, Coen de Graaf & Josep M. Poblet

ICREA, Pg. Lluis Companys 23, Barcelona 08010, Spain

Coen de Graaf

Beijing National Laboratory for Condensed Matter Physics, Institute of Physics, Chinese Academy of Sciences, Beijing, 100190, China

Songshan Lake Materials Laboratory, Dongguan, Guangdong, 523808, China

Heng Gao

Department of Physics & Astronomy, Brigham Young University, Provo, Utah 84602, USA

Branton J. Campbell

CNR-SPIN c/o Università deli Studi dell´Aquila, Via Vetoio 10, I-67100 Coppito (L'Aquila), Italy

Alessandro Stroppa

## Contributions

A.S. conceived and designed the study; A.S., C.G. and J.M.P. supervised the project; F.W. performed the DFT calculations and drafted the manuscript; H.G. constructed the $k \cdot p$ model and assisted the analysis of Rashba anisotropy. B.J.C. conducted the symmetry-mode analysis. All authors reviewed and approved the manuscript.

## Competing interests

The authors declare no competing financial interests.

## Corresponding Authors

Correspondence to Alessandro Stroppa (alessandro.stroppa@spin.cnr.it) or Coen de Graaf (coen.degraaf@urv.cat).

**Supplementary Information**

**for**

**Switchable Rashba Anisotropy in**

**Two-dimensional Hybrid Organic-Inorganic Perovskite**

**by Hybrid Improper Ferroelectricity**

Fei Wang, Heng Gao, Coen de Graaf *, Josep M. Poblet, Branton J. Campbell & Alessandro Stroppa*

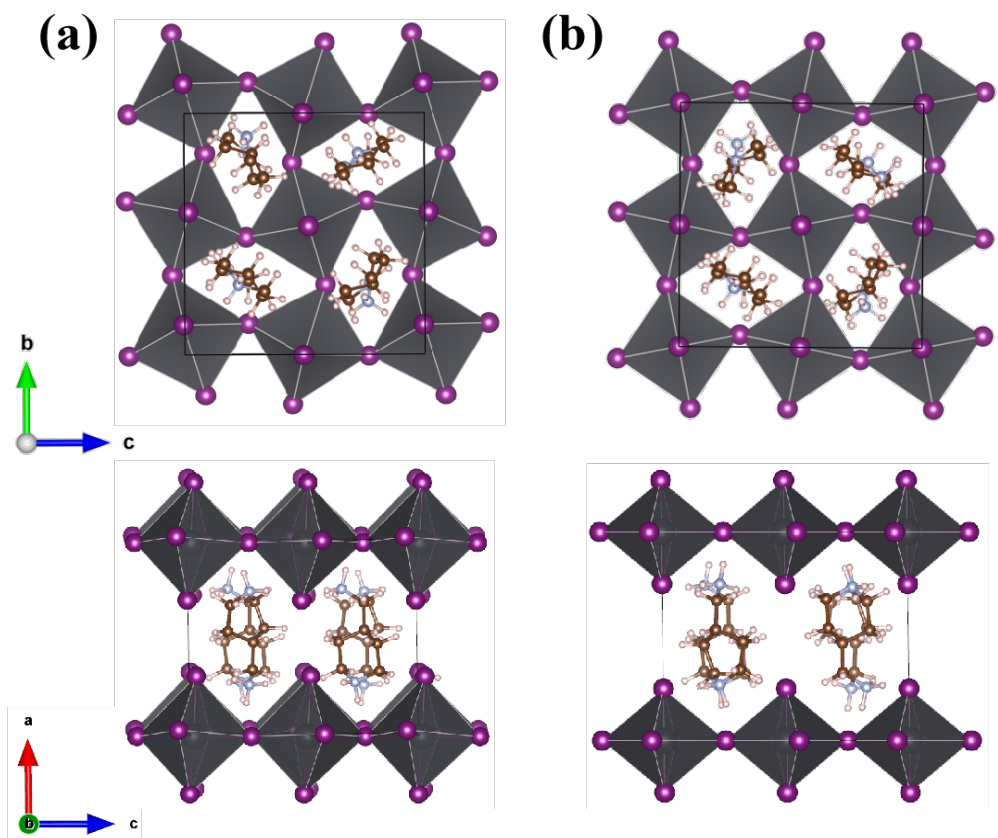

Supplementary Figure 1: **The 1×1×2 supercell of (AMP)PbI$_4$.** **(a)** the ferroelectric structure (FE); **(b)** the anti-ferroelectric structure (AFE). Purple Pb, gray I, brown C, blue N. The protons of the AMP cation were not shown for clarity here.

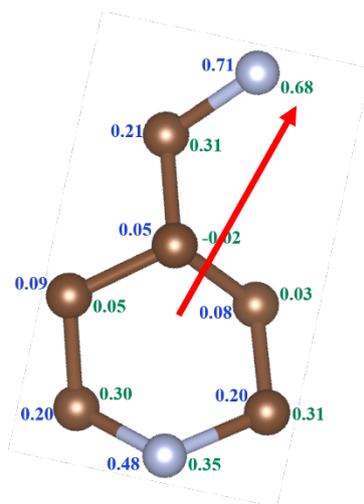

Supplementary Figure 2: **The Hirshfield (blue values) and Natural (green values) population analysis on the isolated AMP cation**, with purple Pb, gray I, brown C and blue N. The charges of hydrogen atoms (omitted) are summed into the heavy atoms. The red arrow indicates the direction of the dipole moment.

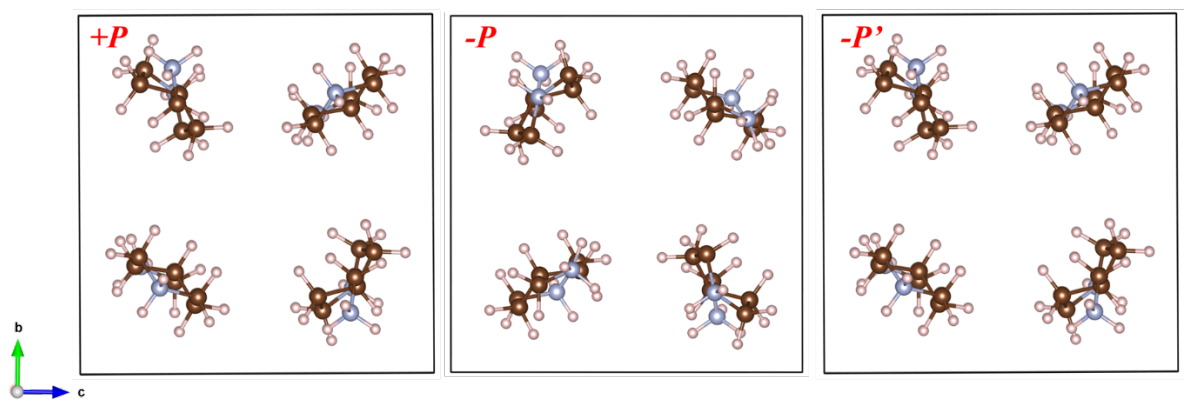

Supplementary Figure 3: **The orientations of four AMP molecules on *bc* plane of +*P*, -*P* and -*P'*,** in which the $PbI_4^{2-}$ inorganic framework is ignore for clarity. Color code: brown C, blue N and pink H.